\documentclass[pdflatex,aps,prl,twocolumn,showpacs,groupedaddress]{revtex4-2}
\usepackage{color,amssymb,amsmath,graphicx,multirow,tabularx,physics}
\usepackage{enumitem}
\usepackage{bm}
\usepackage{tikz}
\usepackage[normalem]{ulem}
\usepackage[caption=false]{subfig}
\usepackage{floatrow}
\definecolor{darkgreen}{rgb}{0.0,0.5,0.0}
\usepackage[colorlinks,citecolor=darkgreen]{hyperref}

\newcommand{\mybraket}[3]{\left[\!\left[ #1 \middle| #2 \middle| #3 \right]\!\right]}
\newcommand{\mybraketb}[2]{\left[\!\left[ #1 \middle| #2 \right]\!\right]}
\newcommand{\mybraketo}[1]{\left[\!\left[ #1 \right]\!\right]}

\begin{document}
\title{
Consistent Field Theory Across the Mott-Insulator to Superfluid Transition
}
\author{Idan S. Wallerstein}
\email{wallersh@post.bgu.ac.il}
\author{Eytan Grosfeld}
\email{grosfeld@bgu.ac.il}
\affiliation{Department of Physics, Ben-Gurion University of the Negev, Beer-Sheva 8410501, Israel}
\begin{abstract}
We employ a field-theoretical approach to analyze the Bose-Hubbard model on a lattice, with a focus on the low-energy properties across the Mott insulator (MI) to superfluid (SF) transition. 
Prior approaches approximated the partition function using cumulant expansions around the MI ground state, which, while accurate in the MI phase, lead to inaccuracies in the SF phase where the MI state is a false ground state.
By expanding around the correct mean-field vacuum, we derive the effective field theory (EFT) governing the MI to SF transition. 
Through this, we reveal the underlying structure of the EFT governing the nucleation of low-energy excitations, particularly the massless and lowest massive modes, offering new insights into their emergence.
\end{abstract}
\maketitle
\emph{Introduction.}--- Quantum field theory is a cornerstone of many-particle physics, providing powerful tools to describe their collective behavior. 
Lattice vibrations in solids, for example, are elegantly captured by mapping their quantum field theoretic description to a collection of harmonic oscillators, representing noninteracting phonons.
The Bose-Hubbard model (BHM), on the other hand, offers a paradigmatic example of an interacting bosonic system that undergoes a quantum phase transition between a Mott insulator (MI) and a superfluid (SF) phase \citep{BHM_Gutzwiller_MF,Intro_ref_1,Intro_ref_3,Intro_ref_4,Intro_ref_5,Intro_ref_6,Intro_ref_8,Intro_ref_9,Intro_ref_10,Intro_ref_11}. 

In the MI phase, the low-lying excitations are either particle-like or hole-like. In the SF phase they can further be classified according to whether they couple to amplitude or phase fluctuations of the order parameter (OP). 
A defining feature is the emergence of a gapless phase-like Goldstone mode, exhibiting linear dispersion at long wavelengths and characterized by a specific speed of sound. 
The SF also supports a gapped amplitude-like Higgs mode. 
These excitations and their decay processes have been the subject of detailed investigation using numerical approaches based on the Gutzwiller ansatz \citep{GMF_quasiparticle_hole_charcteristics,Intro_ref_13,Intro_ref_14,Intro_ref_12,GMF_Quantum_Fluctuations,Intro_ref_19}. 
In the deep SF regime, Bogoliubov mean-field theory serves as a benchmark for the OP and the sound velocity, while in the deep MI regime, the model reduces to exactly solvable independent sites, providing another anchor point for a theory of excitations.

Field-theoretic techniques have played a central role in analyzing the BHM allowing for a physically transparent description in terms of an action. 
The seminal work of Fisher et al. \citep{BHM_phase_diagram_1st} introduced such an effective action via a cumulant expansion following a Hubbard-Stratonovich transformation (HST), yielding reliable results in the MI phase and accurately locating the phase boundary. 
However, the theory quickly loses quantitative accuracy deeper into the SF phase \citep{HST_analytical_approximation,Intro_ref_7}. 
A key development involved applying a second HST \citep{HST_analytical_approximation}, which extends the theory further into the SF regime, though it still falls short of recovering the Bogoliubov limit.

Efforts to derive a low-energy effective field theory (EFT) for the SF phase---built upon these HST-based approaches---inherit the same limitations \citep{Intro_ref_20,Intro_ref_21,Intro_ref_22}. 
As a result, a consistent low-energy EFT of the BHM that captures the correct excitation spectrum and matches the Bogoliubov benchmark remains elusive.

In this paper, we develop a unified framework to describe excitations across both MI and SF phases using a refined field-theoretic approach. Starting from a HST and a cumulant expansion around the mean-field saddle point, we construct an effective local action that decouples into a free theory for the excitations. 
This allows us to systematically characterize both particle-hole modes in the MI and the Goldstone and Higgs modes in the SF. 
At low energies and momenta, our formulation reproduces the gapless mode in the SF and yields an explicit expression for the sound velocity in terms of single-particle Green’s functions. 
Moreover, it provides insights into the nature of the system’s normal modes and the effective field theory that governs them.

\emph{Model and standard field theory.}---
The BHM, introduced in \citep{Bose-Hubbard_1st}, is
governed by the Hamiltonian
\begin{equation}
\label{eq: BHM - Hamiltonian}
    \hat H =  - J\sum\limits_{\left\langle {i,j} \right\rangle }^{} {\hat a_i^\dagger {{\hat a}_j}}  + \frac{U}{2}\sum\limits_i^{} {{{\hat n}_i}\left( {{{\hat n}_i} - 1} \right)}  - \mu \sum\limits_i^{} {{{\hat n}_i}}.
\end{equation}
Here, $i$ labels the lattice sites and $\left\langle {i,j} \right\rangle$ the nearest-neighbor bonds; $\hat a_i^\dagger$ ($\hat a_i$) is the bosonic creation (annihilation) operator and $\hat n_i = \hat a_i^\dagger \hat a_i$ the site occupation number; $J>0$ denotes the tunneling amplitude, $U>0$ the on-site interaction, and $\mu$ the chemical potential.

In the coherent state path integral approach, we study the quantum partition function given by
\begin{align}
  Z &=\int D\left[ {\phi,\phi^* } \right]\exp\left(-S_E[\phi]\right), \quad\text{with,} \\
  S_E[\phi,\phi^*]&=\mybraket{\phi^*}{\partial _\tau  - \mu}{\phi } + \frac{U}{2}\mybraketb{\phi ^{*2}}{\phi ^2} - \mybraket{\phi^*}{J}{\phi}, \nonumber
\end{align}
with $\phi$ the bosonic field, $S_E$ the euclidean action, and
\begin{align}
    \mybraket{\psi}{A}{\phi } \equiv \int\limits_0^\beta  {d{\tau}d{\tau^\prime}\sum\limits_{{i},{j}}^{} {\psi _{{i}}\left( {{\tau}} \right){A_{{i},{j}}}\left( {{\tau},{\tau^\prime}} \right){\phi_{{j}}}\left( {{\tau^\prime}} \right)} }. 
\end{align}
The HST is then used to decouple the sites with the individual sites, retaining the quartic interaction term. 
This comes at the expense of introducing an auxiliary field, $\psi$, which roughly relates to the occupation and phase of the condensate, averaged over the nearest neighbor sites. 
Following integration over $\phi$ and $\phi^*$, the partition function is then written
\begin{align}
  \label{eq:action after HST}
 Z&= \int D\left[ {\psi,\psi^*} \right]\exp(- S_E[\psi,\psi^*]), \quad\text{with,} \\
 S_E[\psi,\psi^*]&={\mybraket{\psi^*}{J^{ - 1}}{\psi} - \ln \left\langle {\exp \left( {\mybraketb{\psi^*}{\phi } + \mybraketb{\phi^*}{\psi}} \right)} \right\rangle_\text{loc}}. \nonumber
\end{align}
Here $\langle\ldots\rangle_\text{loc}$ is the thermal average taken with respect to $\hat H_{\text{loc}}$ comprising the second and third terms in Eq.~(\ref{eq: BHM - Hamiltonian}).

\begin{figure}
    \centering
    \includegraphics[width=1\linewidth]{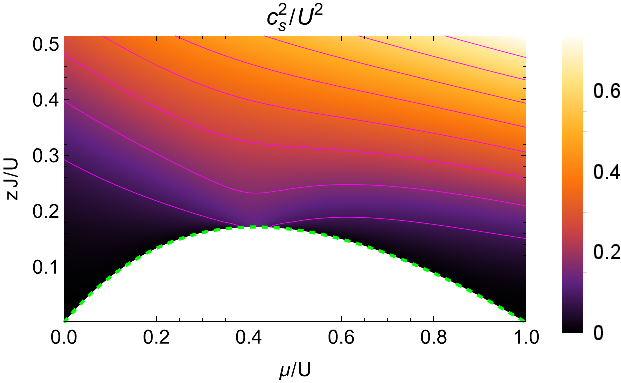}
    \caption{The dimensionless sound velocity $c_s^2/U^2$ as function $\mu$ and $J$ within the SF phase.
    The dashed green line indicates the boundary of the SF with the $n_0=1$ Mott lobe; within the latter, the sound velocity is ill-defined (white).
    Contour lines are colored in Magenta. We take the lattice constant $\text{a}=1$, for the 1D chain.}
    \label{fig: phase diagram}
\end{figure}

Due to the interactions, the average cannot be evaluated exactly, but instead is approximated in the standard approach \citep{BHM_phase_diagram_1st}, using the cumulant expansion.
The resulting Landau-like action is a functional of $\psi$ and $\psi^*$, which is number-conserving due to the U$(1)$ symmetry of the local Hamiltonian.
A quantum phase transition is found between the MI phase and the SF phase, which for a constant homogeneous real field $\psi_i(\tau)=\psi$, is marked by a change in sign of the coefficient of $\left|\psi\right|^2$ \citep{HST_analytical_approximation}. 
The different phases are characterized by the OP given by the expectancy $\langle\psi\rangle_\text{loc}$. 
The cumulant expansion gives the same phase diagram (Fig.~\ref{fig: phase diagram}) as the standard mean-field or the Gutzwiller mean field (GMF) approximations \citep{BHM_Gutzwiller_MF}. 
The expansion predicts a vanishing OP in the MI phase, as required.
In contrast, in the SF phase, the cumulant expansion predicts a vanishing OP as $J/U\to\infty$ \cite{author2025supplement}, which contradicts the Bogolyubov approximation of a weakly-interacting Bose gas \citep{Bogolyubov_spectrum}, and leads to an unphysical excitation spectrum~\citep{HST_Cumulant_is_bad}.

\emph{Derivation of the EFT.}--- The failure of the cumulant expansion is due to expanding around a local maximum at $\psi=0$.
In the SF phase, the OP acquires a finite value around which the auxiliary field should be expanded, that is, taking $\psi=\psi_{\text{min}}+\delta\psi$.
We take $\psi_{\text{min}}$ to be the constant homogeneous solution to the saddle-point equation, for the relevant action Eq.~(\ref{eq:action after HST}).
This saddle-point approximation is equivalent to the Gutzwiller's mean field (GMF) approach \citep{HST_MF_equivalence2}, thus the minimizing value is the GMF OP, $\psi_{\text{min}}=\psi^{(0)}$, which due to the $U(1)$ symmetry of the BHM, we choose to be a pure real number.
The evaluation of $\psi^{(0)}$ is performed numerically, therefore, it requires a truncation of the Fock space, i.e., a maximal occupation number $n_{\text{max}}$ for each site~\citep{BHM_Gutzwiller_MF}.

Next, we study the deviations from the GMF approximation, $\delta\psi$, i.e., the excitations,
\begin{align}
 &Z= \int D\left[ {\delta\psi,\delta\psi^*} \right]\exp(- S_E[\delta\psi,\delta\psi^*]), \quad\text{with,} \\
&S_E[\delta\psi,\delta\psi^*]=\mybraket{\delta\psi^*}{J^{ - 1}}{\psi^{(0)}} + \mybraket{\psi^{(0)}}{J^{ - 1}}{\delta\psi} \nonumber \\
 &+ \mybraket{\delta\psi^*}{J^{ - 1}}{\delta\psi} - \ln {{\left\langle {\exp \left( {\mybraketb{\delta\psi^*}{\phi} + \mybraketb{\phi^*}{\delta\psi}} \right)} \right\rangle }_\text{MF}}. \nonumber
\end{align}
Here $\langle\ldots\rangle_\text{MF}$ is the thermal average with respect to a $\psi^{(0)}$-dependent modified mean-field Hamiltonian
\begin{align}
\label{eq: BHM - Mean Field Hamiltonian}
    {{\hat H}_{\text{MF},i}} = \frac{U}{2}{{\hat n}_i}\left( {{{\hat n}_i} - 1} \right) - \mu {{\hat n}_i} - \psi^{(0)} \left(\hat a_i^\dagger + \hat a_i\right).
\end{align}
Eq.~(\ref{eq: BHM - Mean Field Hamiltonian}) is not number-conserving. 
The action for the excitations is then to quadratic order in the cumulant expansion
\begin{align}
  &S\left[ {{\delta\psi,\delta\psi^*}} \right]\approx \mybraket{\delta\psi^*}{J^{ - 1} - \left\langle {{{\left| {\delta \phi } \right|}^2}} \right\rangle_\text{MF}}{\delta\psi} \\
 &- \frac{1}{2}\mybraket{\delta\psi^*}{\left\langle {\delta {\phi ^2}} \right\rangle_\text{MF}}{\delta\psi^*} - \frac{1}{2}\mybraket{\delta\psi}{{\left\langle {\delta {\phi ^{*2} }} \right\rangle_\text{MF} }}{\delta\psi} , \nonumber
\end{align}
where $\delta\phi=\phi-\langle\phi\rangle_\text{MF}$ is the fluctuation of the original Bosonic field around the average $\langle\phi\rangle_\text{MF}=\langle\hat{a}_i\rangle_\text{MF}=\left(zJ\right)^{-1}\psi^{(0)}$, with $z$ the coordination number. 
The overall linear term vanishes due to cancelation of the explicit linear terms in the action and the linear terms in the cumulant expansion, as expected from the expansion of the action around its minimum value.

Writing the action using Green functions in Fourier space, we have
\begin{align}
\label{eq: BHM - GMF+HST - action}
S\left[ {\delta \psi,\delta\psi^* } \right] = \frac{1}{2}\sum\limits_{{\mathbf{k}},\omega }^{} {\delta \Psi _{\mathbf{k}}^\dagger \left( {i\omega } \right){\mathcal{G}}\left( {{\mathbf{k}},i\omega } \right)\delta \Psi _{\mathbf{k}}\left( {i\omega } \right)} ,
\end{align}
with
\begin{align}
    \delta \Psi^\dagger_{\mathbf{k}}\left( {i\omega } \right)&=\left(
  {\delta \psi _{\mathbf{k}}^*\left( {i\omega } \right)},{\delta \psi _{ - {\mathbf{k}}}^{}\left( { - i\omega } \right)}\right) \quad \text{and}\nonumber\\
    {\mathcal{G}}\left( {{\mathbf{k}},i\omega } \right)&={\left( {\begin{array}{*{20}{c}}
  {J_{\mathbf{k}}^{ - 1} + G\left( {i\omega } \right)}&{{G^\prime }\left( {i\omega } \right)} \\ 
  {{G^\prime }\left( { - i\omega } \right)}&{J_{ - {\mathbf{k}}}^{ - 1} + G\left( { - i\omega } \right)} 
\end{array}} \right)} ,
\end{align}
where $J_{\mathbf{k}}$ is the spatial Fourier transform of the tunnelling matrix, while $G(i\omega)$ and $G^\prime(i\omega)$ are the Fourier transformed Green function, $G_{i,j}(\tau-\tau^\prime)=-\langle\delta\phi_i(\tau)\delta\phi_j^*(\tau^\prime)\rangle_\text{MF}$ and the anomalous Green function, $G^\prime_{i,j}(\tau-\tau^\prime)=-\langle\delta\phi_i(\tau)\delta\phi_j(\tau^\prime)\rangle_\text{MF}=-\langle\delta\phi_i^*(\tau)\delta\phi_j^*(\tau^\prime)\rangle_\text{MF}$, respectively, where the last equality is due to the $\phi\to\phi^*$ symmetry in Eq.~(\ref{eq: BHM - Mean Field Hamiltonian}) resulted from the gauge choice of real $\psi^{(0)}$.
The time dependence of the anomalous Green function is a necessary component to arrive at the correct excitation spectrum in the SF phase \citep{Popov_book}.

We find the spectrum as the on-shell frequencies of the quadratic action Eq.~(\ref{eq: BHM - GMF+HST - action}) after analytic continuation $i\omega\to\omega$.
Meaning, the spectrum equation is given by a vanishing determinant of the Green matrix $\mathcal{G}$, yielding $\omega_{\lambda,{\mathbf{k}}}$ with $\lambda$ indexing the excitation bands, while the corresponding kernel, $(\mathcal{U}_{\lambda,{\mathbf{k}}},\mathcal{V}_{\lambda,{\mathbf{k}}}^*)^\intercal$, comprises the normalized quasi- particle and hole amplitudes that characterize the excitations \citep{GMF_quasiparticle_hole_charcteristics}.
Equivalently, the excitation spectrum is extracted from the poles of the one-particle correlator $\langle\delta\psi_{\mathbf{k}}\left( \omega  \right)\delta\psi_{\mathbf{k}}^*\left( \omega  \right)\rangle_\text{MF}$.
In the MI phase, the anomalous Green function vanishes due to the U$(1)$ symmetry of the MF Hamiltonian, thus the Green matrix is
\begin{align}
    \nonumber & \mathcal{G}_\text{MI}({\mathbf{k}},\omega)=\left( {\begin{array}{*{20}{c}}
  {J_{\mathbf{k}}^{ - 1} + G_\text{MI}\left( \omega  \right)}&0 \\ 
  0&{J_{ - {\mathbf{k}}}^{ - 1} + G_\text{MI}\left( { - \omega } \right)} 
\end{array}} \right), \text{with}, \\
    & G_\text{MI}\left( \omega  \right) = -\frac{{{n_0}}}{{\omega  + \mu  - U\left( {{n_0} - 1} \right)}} + \frac{{{n_0} + 1}}{{\omega  + \mu  - U{n_0}}},
\end{align}
and ${n_0} = \left\lceil {\max \left( {0,\mu /U} \right)} \right\rceil$ is the occupation number in the corresponding Mott lobe in the phase diagram.
The spectrum is composed of the two lowest gapped bands found in the GMF.
The corresponding kernel is given such that one of the bands $\lambda,\lambda'$ is of pure particle excitations ($\mathcal{U}_{\lambda,{\mathbf{k}}}=\delta\psi_{\mathbf{k}}(\omega_{\lambda,{\mathbf{k}}})=1$, $\mathcal{V}_{\lambda,{\mathbf{k}}}=\delta\psi_{-{\mathbf{k}}}(-\omega_{\lambda,{\mathbf{k}}})=0$), while the other is of pure hole ($\mathcal{U}_{\lambda^\prime,{\mathbf{k}}}=\delta\psi_{\mathbf{k}}(\omega_{\lambda^\prime,{\mathbf{k}}})=0$, $\mathcal{V}_{\lambda^\prime,{\mathbf{k}}}=\delta\psi_{-{\mathbf{k}}}(-\omega_{\lambda^\prime,{\mathbf{k}}})=1$), depending on the chemical potential.
The GMF approximation, in addition to the two gapped bands, finds higher multi-particle flat bands, which are absent in the field theory approach as they are not described by the one-particle correlator $\langle\delta\psi_{\mathbf{k}}\left( \omega  \right)\delta\psi_{\mathbf{k}}^*\left( \omega  \right)\rangle_\text{MI}$.

\begin{figure}
    \centering
    \includegraphics[width=0.99\linewidth]{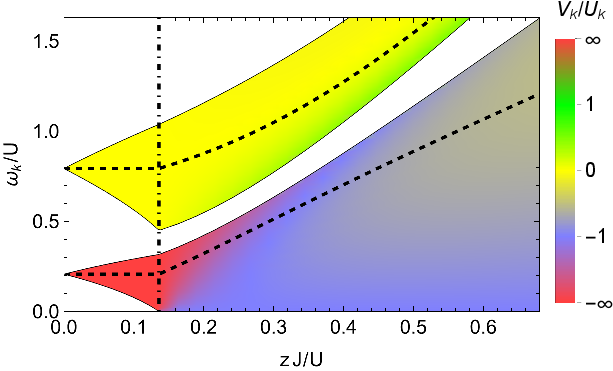}
    \caption{Excitations spectrum $\omega_{\lambda,k}$ as a function of $J/U$,
    showing the two lowest bands ($\lambda=1,2$), corresponding singularities, $\omega_{\lambda,\pi/2}$ (black dashed lines) and phase transition point (vertical black dot-dashed).
    Color indicates the ratio between particle and hole amplitudes of the excitation:
    pure hole (red), pure particle (yellow), pure phase (blue), and pure amplitude excitation (green).
    In the MI phase, the excitations are pure particle/hole, while in the SF near the phase transition, the excitation inherits the MI characteristics.
    The $k\approx0$ excitations are mostly phase and amplitude for the gapless and gapped band, respectively, where pure phase and amplitude excitations are found at points of a relativistic effective theory (Fig.~\ref{fig:magnetic field}).
    For larger $J/U$, the gapless band becomes particle like, as predicted by the Bogolyubov spectrum \citep{Bogolyubov_spectrum}. 
    Here $\mu/U=  (\mu/U)_c/2$, where $(\mu/U)_c$ is the value for the tip of the Mott lobe, for the 1D chain.
    }
    \label{fig:Excitation spectrum - U/V color}
\end{figure}

In the SF phase, the spectrum is evaluated numerically and both $G$ and $G^\prime$ are sums of $2n_{\text{max}}$ rational functions, leading to a polynomial equation of degree $n_{\text{max}}$ for the variable $\omega^2$. 
The resulting spectrum is $n_{\text{max}}$ bands identical to the spectrum found in the GMF approximation (Fig.~\ref{fig:Excitation spectrum - U/V color}) \citep{BHM_Gutzwiller_MF}.
Each band, due to the anomalous Green function, includes excitations that are a mix of particle-hole ($0\le \left|\mathcal{U}_{\mathbf{k}}\right|,\left|\mathcal{V}_{\mathbf{k}}\right|\le1$) \citep{GMF_quasiparticle_hole_charcteristics}.

An alternative approach to get the excitations is to use the diagonalized Green matrix
\begin{align}
    {\mathcal{G}}\left( {{\mathbf{k}},i\omega } \right)&={\left( {\begin{array}{*{20}{c}}
  { G_+\left({\mathbf{k}}, {i\omega } \right)}&{0} \\ 
  {0}&{G_-\left( {-{\mathbf{k}}, - i\omega } \right)} 
\end{array}} \right)} ,
\end{align}
in the basis of the new fields $\left( {\varphi_{+,{\mathbf{k}}}\left( {i\omega } \right)},{\varphi_{-,{\mathbf{k}}}\left( {i\omega }\right)}\right)^\intercal$.
The equations of the excitation spectrum are now $G_\pm({\mathbf{k}},\omega)=0$, each gives $n_{\max}/2$ bands with the kernel of $\left(1,0\right)^\intercal$ or $\left(0,1\right)^\intercal$, and the particle/hole amplitudes can be evaluated using the diagonalizing matrix.

\emph{Low energy EFT.}--- The expansion of the Green functions to low momenta and energy gives the experimentally accessible excitations, mainly the gapless mode.
Denoting ${G}({\mathbf{k}},i\omega)=J_{\mathbf{k}}^{ - 1} + G\left( i\omega  \right)$, the Green matrix is expanded to give
\begin{align}
  \mathcal{G}({\mathbf{k}},i\omega)\approx \left( {\begin{array}{*{20}{c}}
  {A({\mathbf{k}},i\omega)}&{B(i\omega)} \\ 
  {B(-i\omega)}&{A(-{\mathbf{k}},-i\omega)} 
\end{array}} \right) ,
\end{align}
with
\begin{align}
    A({\mathbf{k}},i\omega)&={G} + \frac{{{k^2}}}{2}{{G}^{\left( {2,0} \right)}} + i\omega {{G}^{\left( {0,1} \right)}} - \frac{{{\omega ^2}}}{2}{{G}^{\left( {0,2} \right)}} \quad \text{and} \nonumber\\
    B(i\omega)&={G}' + i\omega {{{G}'}^{\left( 1 \right)}} - \frac{{{\omega ^2}}}{2}{{G}'}^{\left( 2 \right)} ,
\end{align}
where the coefficients ${G}, {G^\prime}$  and their derivatives w.r.t. ${\mathbf{k}}$ and $i\omega$ are evaluated at $k=0,i\omega=0$ and are real.
We can take $G^{\prime(1)}=0$ as it is evaluated numerically to vanish throughout the phase diagram in addition to being a coefficient of a full derivative term in the time domain action.

In the MI phase, the Green matrix is already diagonalized, and the effective action is simply
\begin{equation}
    {S^{\left( {{\text{MI}}} \right)}}\left[ {{\varphi _ + },{\varphi _ - }} \right] = \frac{1}{2}\sum\limits_{{\mathbf{k}},\omega ,\sigma  =  \pm }^{} {A\left( {\sigma {\mathbf{k}},\sigma i\omega } \right){{\left| {{\varphi _{\sigma ,{\mathbf{k}}}}\left( {i\omega } \right)} \right|}^2}} .
\end{equation}
The approximated spectrum is of the form ${\omega _ \pm } \approx \Delta _ \pm ^{\left( {{\text{MI}}} \right)} + k^2/2 m^{\left( {{\text{MI}}} \right)}$, where $\Delta^{\text{(MI)}}_\pm$ and $m^{\text{(MI)}}$ are given in \cite{author2025supplement}.
In the SF phase, the action can be cast in the form
\begin{align}
\label{eq: excitation action - by band fields}
  &{S^{\left( {{\text{SF}}} \right)}}\left[ {{\varphi _ + },{\varphi _ - }} \right] = \\ &\frac{{{\rho_s^{(1)}}}}{2}\sum\limits_{{\mathbf{k}},\omega ,\sigma  =  \pm }^{} {\left[ {{{\left( {{m_\sigma^{\text{(SF)}} }{c_\sigma }} \right)}^2} + {k^2} + c_\sigma ^{ - 2}{\omega ^2}} \right]{{\left| {{\varphi _{\sigma ,{\mathbf{k}}}}\left( {i\omega } \right)} \right|}^2}}. \nonumber
\end{align}
It represents a Klein-Gordon action, with mass $m_\pm^{\text{(SF)}}\propto \sqrt{G\pm G^\prime}$ (other coefficients are given explicitly in \cite{author2025supplement}).

The spectrum becomes gapless in the SF phase, as expected by the Goldstone theorem. 
The mode $\varphi_-$ acquires a zero mass ($G=G^\prime$) throughout the SF phase, as is also verified numerically.
The low-energy band is thus governed by the hydrodynamical action
\begin{equation}
\label{eq:Goldstone action}
    S_\text{HD}\left[\varphi_-\left({\bf r},\tau\right)\right]=\int{d\tau d{\mathbf{r}}\frac{\rho^{(1)}_s}{2}\left[{{\left( {\nabla \varphi_-} \right)}^2} + c_s^{-2} {{\left( {\partial_\tau \varphi_-} \right)}^2}\right]},
\end{equation}
where the coefficient $\rho^{(1)}_s={{G^{\left( {2,0} \right)}}}/2$ characterizes the stiffness of the low energy single-particle excitation, rather than the full superfluid density, which is determined by the response of the action to a local twist field $\theta(\bf{r},\tau)$ \citep{Bog_superfluid_fraction}.
The speed of sound $c_s$ characterizing the dispersion relation $\omega_-=c_sk$ for low momenta is
\begin{equation}
\label{eq: speed of sound}
    c_-^{2}=c_s^{2}=\frac{{G{G^{\left( {2,0} \right)}}}}{{{{\left( {{G^{\left( {0,1} \right)}}} \right)}^2} - G\left( {{G^{\left( {0,2} \right)}} - {{G}^{\prime\left( 2 \right)}}} \right)}} .
\end{equation}
This expression is numerically identical to the GMF results (Fig.~\ref{fig: phase diagram}) \citep{BHM_Gutzwiller_MF}.

The mode $\varphi_+$, described by a massive (gapped) Klein-Gordon action, admits a quadratic dispersion relation for small $k$, $\omega_+=\Delta_+^{\text{(SF)}}+k^2/2m_+^{\text{(SF)}}$.
The band gap is given in term of the Green functions by
\begin{align}
\label{eq: upper band gap}
    \Delta _ + ^{\left( {{\text{SF}}} \right)} = m_ + ^{\left( {{\text{SF}}} \right)}c_ + ^{\text{2}} = \sqrt {\frac{{ - 4{G^2}}}{{{{\left( {{G^{(0,1)}}} \right)}^2} + G\left( {{G^{(0,2)}} + {{G'}^{\left( 2 \right)}}} \right)}}} .
\end{align}

\begin{figure}
    \centering
    \includegraphics[width=1\linewidth]{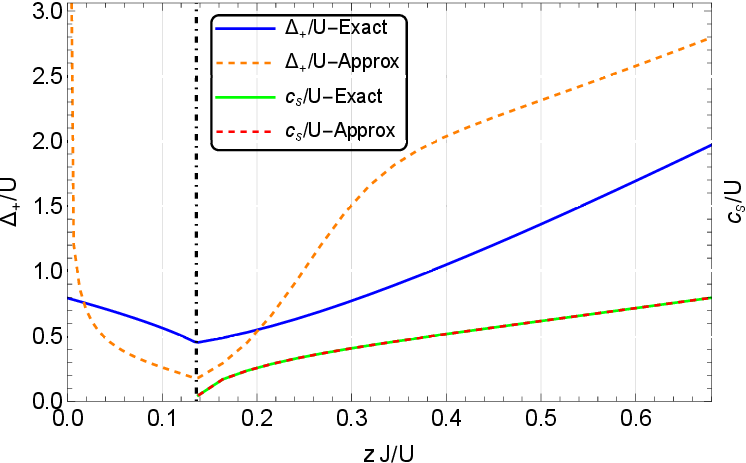}
    \caption{Speed of sound of the gapless mode $c_s$ and the gap of the upper band $\Delta_+=\omega_{+,k=0}$ calculated using the full Green matrix and using the low energy effective theory (Eq.~\ref{eq: speed of sound} and Eq.~\ref{eq: upper band gap}).
    The discrepancy in the gap but not in the speed of sound is due to the gap being outside of the low-energy EFT radius of convergence (Fig.~\ref{fig:Excitation spectrum - U/V color}).
    Vertical black dot-dashed line indicates the phase transition point. Here $\mu/U=(\mu/U)_c/2$.}
    \label{fig:Gap comparison}
\end{figure}

We note that the low-frequency expansion accurately approximates the speed of sound (Fig.~\ref{fig:Gap comparison}), but the band gap $\Delta_+$ agrees only qualitatively with the excitation gap as determined by GMF (Fig.~\ref{fig:Gap comparison}). 
The reason behind this loss of accuracy is in the expansion of the singular Green functions in $\omega$. 
This expansion has a finite radius of convergence, given by the nearest pole to the point of expansion, $\omega=0$.
The poles $\omega_\lambda({\mathbf{k}}_r)$ of the Green matrix determinant are located at $J_{{\mathbf{k}}_r}=0$ for each band, see Fig.~\ref{fig:Excitation spectrum - U/V color}. 
The value of the lowest pole separates the lower band from the first gapped band, hence the latter is not accurately described by the low-energy EFT.

\floatsetup[figure]{style=plain,subcapbesideposition=top}
\begin{figure}
\sidesubfloat[]{ 
\label{fig:spring constant}
\includegraphics[width=1\linewidth]
{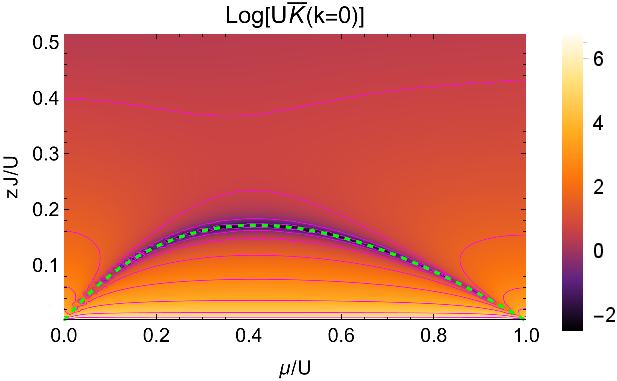}} \hfil
\sidesubfloat[]{
\label{fig:magnetic field}
\includegraphics[width=1\linewidth]
{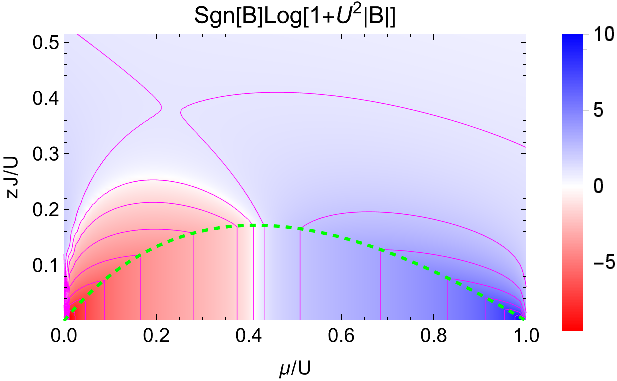}}
\caption{
Dimensionless parameters for the low energy effective Hamiltonian: 
(a) Average spring constant $2\bar{K}=K_h+K_g=4G$, with the phase boundary (dashed green) coinciding with the $\bar{K}=0$ line; and, (b) Uniform magnetic field strength $B=2G^{(0,1)}$.
A vanishing magnetic field corresponds to a particle-hole symmetry, $\left|\mathcal{U}_k\right|=\left|\mathcal{V}_k\right|$, in both the gapless and lowest gapped bands, corresponding to pure amplitude (Higgs) and pure phase (Goldstone) modes in a relativistic effective theory. 
Contour lines depicted in magenta.}
\end{figure}

\emph{Dynamics of pure amplitude and phase excitations.}
For a real OP, the pure amplitude and the pure phase excitations are along the real ($h$) and imaginary ($g$) parts of the OP perturbation,  $\delta\psi\left({\bf r},\tau\right)=h\left({\bf r},\tau\right)+ig\left({\bf r},\tau\right)$.
They can also be considered as a superposition of particle and hole excitations or, alternatively, of upper- and lower- band excitations. 
The corresponding Lagrangian density is given by
\begin{align}
\label{eq:LEEFT}
    &\mathcal{L}\left[ {h\left( {{\mathbf{r}},\tau } \right),g\left( {{\mathbf{r}},\tau } \right)} \right] = \left( {G + G'} \right){h^2} + \left( {G - G'} \right){g^2} \hfill \\
    &+ \frac{1}{2}{G^{\left( {2,0} \right)}}\left( {{{\left( {\nabla h} \right)}^2} + {{\left( {\nabla g} \right)}^2}} \right) + i{G^{\left( {0,1} \right)}}\left( {g{\partial _\tau }h - h{\partial _\tau }g} \right) \hfill \nonumber\\
   &- \frac{1}{2}{\left( {{\partial _\tau }h} \right)^2}\left( {{G^{\left( {0,2} \right)}} + {{G'}^{\left( 2 \right)}}} \right) - \frac{1}{2}{\left( {{\partial _\tau }g} \right)^2}\left( {{G^{\left( {0,2} \right)}} - {{G'}^{\left( 2 \right)}}} \right) . \nonumber
\end{align}
Eq.~(\ref{eq:LEEFT}) encapsulates the quadratic action in terms of the Green functions and their derivatives across the SF-to-MI transition.
To study the dynamics in a familiar form, we write the Hamiltonian given by the Legendre transformation of the real-time Lagrangian ($\mathcal{L}(\tau)\to-\mathcal{L}(i t)$),
\begin{align}
\label{eq:LEFH}
    {{\mathcal{H}}^+_{\mathbf{k}}} &= \frac{{{{\left| {{p_{h,{\mathbf{k}}}} + B{g_{\mathbf{k}}}/2} \right|}^2}}}{2M_h} +\frac{K_h({\mathbf{k}}){\left| {{h_{\mathbf{k}}}} \right|^2}}{2} \nonumber\\
    & + \frac{{{{\left| {{p_{g,{\mathbf{k}}}} - B{h_{\mathbf{k}}}/2} \right|}^2}}}{2M_g} + \frac{K_g({\mathbf{k}}){\left| {{g_{\mathbf{k}}}} \right|^2}}{2} , 
\end{align}
where ${\mathcal{H}}^+_{\mathbf{k}}$ is the Hamiltonian for real time $t$ and for half of the $k$-space. 
The fields $h_{\mathbf{k}}(t)$ and $g_{\mathbf{k}}(t)$ are the spatial Fourier transformed complex fields of $h$ and $g$, respectively, and $p_{h,{\mathbf{k}}}(t)$ and $p_{g,{\mathbf{k}}}(t)$ are the corresponding canonical momenta.

In the MI phase, where the anomalous Green function vanishes, the masses are equal, $M_h=M_g$ as well as the spring constants, $K_h=K_g$, hence Eq.~(\ref{eq:LEFH}) describes an isotropic two-dimensional harmonic oscillator with a $k$ dependent spring constant and a constant magnetic field.
In the SF phase, Eq.~(\ref{eq:LEFH}) describes an anisotropic harmonic oscillator in a constant magnetic field. The expressions for the parameters as functions of the Green functions are in \citep{author2025supplement}.

The low-energy approximation for the excitations spectrum is given by the frequencies of the normal modes of the Hamiltonian, i.e., the Hamiltonian is written in the form $\mathcal{H}^+_{\mathbf{k}}=\omega_{+,{\mathbf{k}}}a^*_{+,{\mathbf{k}}}a_{+,{\mathbf{k}}}+\omega_{-,{\mathbf{k}}}a^*_{-,{\mathbf{k}}}a_{-,{\mathbf{k}}}$ \citep{Harmonic_Solution}.  
These frequencies are functions of the oscillator and cyclotron frequencies as expressed by the Hamiltonian parameters.
The spring constants at $k=0$, $K_{h,g}=2(G\pm G^\prime)$ vanish at the phase transition, when the lower band gap closes (Fig.~\ref{fig:spring constant}), while a finite magnetic field $B=2G^{(0,1)}$ breaks the particle-hole symmetry, $\left|\mathcal{U}_{\mathbf{k}}\right|\ne\left|\mathcal{V}_{\mathbf{k}}\right|$, and leads to a non-relativistic effective theory. 
The magnetic field also sets the gap between the lower and upper band in the MI phase (Fig.~\ref{fig:magnetic field}).

\emph{Conclusions.}--- We have developed a field-theoretical framework that captures quantum fluctuations across the entire MI to SF transition. 
In both phases, the method yields a local effective action: number-conserving in the MI phase and non-conserving in the SF phase, consistent with the emergence of spontaneous symmetry breaking. 
The accuracy of the excitation spectrum in the SF phase hinges on treating the anomalous Green's function as a dynamical quantity.

Compared to GMF approaches, our field-theoretical description allows for a simple interpretation of the excitations and their dynamics by mapping the effective action onto a set of independent two-dimensional harmonic oscillators, each with lattice-momentum-dependent anisotropic harmonic potential, influenced by a perpendicular faux-magnetic field. 
The interplay between these elements gives rise to gapless traditional phonons---the hallmark of the broken symmetry phase---and to the gapped amplitude mode. 
The location of the phase transition line is marked by the vanishing of the harmonic potential at zero lattice momentum, where the gap to the amplitude mode is set by the cyclotron frequency associated with the faux-magnetic field.

Compared to previous field-theoretical descriptions, our unified approach treats the two phases on an equal theoretical footing, enabling a consistent description across the quantum critical point. Importantly, the speed of sound agrees with the prediction of Bogoliubov.

\begin{acknowledgments}
    We thank D. Cohen for insightful discussions.
\end{acknowledgments}

\bibliography{refs.bib}

\begin{thebibliography}{30}%
\makeatletter
\providecommand \@ifxundefined [1]{%
 \@ifx{#1\undefined}
}%
\providecommand \@ifnum [1]{%
 \ifnum #1\expandafter \@firstoftwo
 \else \expandafter \@secondoftwo
 \fi
}%
\providecommand \@ifx [1]{%
 \ifx #1\expandafter \@firstoftwo
 \else \expandafter \@secondoftwo
 \fi
}%
\providecommand \natexlab [1]{#1}%
\providecommand \enquote  [1]{``#1''}%
\providecommand \bibnamefont  [1]{#1}%
\providecommand \bibfnamefont [1]{#1}%
\providecommand \citenamefont [1]{#1}%
\providecommand \href@noop [0]{\@secondoftwo}%
\providecommand \href [0]{\begingroup \@sanitize@url \@href}%
\providecommand \@href[1]{\@@startlink{#1}\@@href}%
\providecommand \@@href[1]{\endgroup#1\@@endlink}%
\providecommand \@sanitize@url [0]{\catcode `\\12\catcode `\$12\catcode
  `\&12\catcode `\#12\catcode `\^12\catcode `\_12\catcode `\%12\relax}%
\providecommand \@@startlink[1]{}%
\providecommand \@@endlink[0]{}%
\providecommand \url  [0]{\begingroup\@sanitize@url \@url }%
\providecommand \@url [1]{\endgroup\@href {#1}{\urlprefix }}%
\providecommand \urlprefix  [0]{URL }%
\providecommand \Eprint [0]{\href }%
\providecommand \doibase [0]{https://doi.org/}%
\providecommand \selectlanguage [0]{\@gobble}%
\providecommand \bibinfo  [0]{\@secondoftwo}%
\providecommand \bibfield  [0]{\@secondoftwo}%
\providecommand \translation [1]{[#1]}%
\providecommand \BibitemOpen [0]{}%
\providecommand \bibitemStop [0]{}%
\providecommand \bibitemNoStop [0]{.\EOS\space}%
\providecommand \EOS [0]{\spacefactor3000\relax}%
\providecommand \BibitemShut  [1]{\csname bibitem#1\endcsname}%
\let\auto@bib@innerbib\@empty
\bibitem [{\citenamefont {Krutitsky}\ and\ \citenamefont
  {Navez}(2011)}]{BHM_Gutzwiller_MF}%
  \BibitemOpen
  \bibfield  {author} {\bibinfo {author} {\bibfnamefont {K.~V.}\ \bibnamefont
  {Krutitsky}}\ and\ \bibinfo {author} {\bibfnamefont {P.}~\bibnamefont
  {Navez}},\ }\href {https://doi.org/10.1103/PhysRevA.84.033602} {\bibfield
  {journal} {\bibinfo  {journal} {Phys. Rev. A}\ }\textbf {\bibinfo {volume}
  {84}},\ \bibinfo {pages} {033602} (\bibinfo {year} {2011})}\BibitemShut
  {NoStop}%
\bibitem [{\citenamefont {Sajna}\ \emph {et~al.}(2015)\citenamefont {Sajna},
  \citenamefont {Polak}, \citenamefont {Micnas},\ and\ \citenamefont
  {Ro\ifmmode~\dot{z}\else \.{z}\fi{}ek}}]{Intro_ref_1}%
  \BibitemOpen
  \bibfield  {author} {\bibinfo {author} {\bibfnamefont {A.~S.}\ \bibnamefont
  {Sajna}}, \bibinfo {author} {\bibfnamefont {T.~P.}\ \bibnamefont {Polak}},
  \bibinfo {author} {\bibfnamefont {R.}~\bibnamefont {Micnas}},\ and\ \bibinfo
  {author} {\bibfnamefont {P.}~\bibnamefont {Ro\ifmmode~\dot{z}\else
  \.{z}\fi{}ek}},\ }\href {https://doi.org/10.1103/PhysRevA.92.013602}
  {\bibfield  {journal} {\bibinfo  {journal} {Phys. Rev. A}\ }\textbf {\bibinfo
  {volume} {92}},\ \bibinfo {pages} {013602} (\bibinfo {year}
  {2015})}\BibitemShut {NoStop}%
\bibitem [{\citenamefont {L\"uhmann}(2013)}]{Intro_ref_3}%
  \BibitemOpen
  \bibfield  {author} {\bibinfo {author} {\bibfnamefont {D.-S.}\ \bibnamefont
  {L\"uhmann}},\ }\href {https://doi.org/10.1103/PhysRevA.87.043619} {\bibfield
   {journal} {\bibinfo  {journal} {Phys. Rev. A}\ }\textbf {\bibinfo {volume}
  {87}},\ \bibinfo {pages} {043619} (\bibinfo {year} {2013})}\BibitemShut
  {NoStop}%
\bibitem [{\citenamefont {Amico}\ and\ \citenamefont
  {Penna}(1998)}]{Intro_ref_4}%
  \BibitemOpen
  \bibfield  {author} {\bibinfo {author} {\bibfnamefont {L.}~\bibnamefont
  {Amico}}\ and\ \bibinfo {author} {\bibfnamefont {V.}~\bibnamefont {Penna}},\
  }\href {https://doi.org/10.1103/PhysRevLett.80.2189} {\bibfield  {journal}
  {\bibinfo  {journal} {Phys. Rev. Lett.}\ }\textbf {\bibinfo {volume} {80}},\
  \bibinfo {pages} {2189} (\bibinfo {year} {1998})}\BibitemShut {NoStop}%
\bibitem [{\citenamefont {Wang}\ \emph {et~al.}(2018)\citenamefont {Wang},
  \citenamefont {Zhang}, \citenamefont {Hou}, \citenamefont {Eggert},\ and\
  \citenamefont {Pelster}}]{Intro_ref_5}%
  \BibitemOpen
  \bibfield  {author} {\bibinfo {author} {\bibfnamefont {T.}~\bibnamefont
  {Wang}}, \bibinfo {author} {\bibfnamefont {X.-F.}\ \bibnamefont {Zhang}},
  \bibinfo {author} {\bibfnamefont {C.-F.}\ \bibnamefont {Hou}}, \bibinfo
  {author} {\bibfnamefont {S.}~\bibnamefont {Eggert}},\ and\ \bibinfo {author}
  {\bibfnamefont {A.}~\bibnamefont {Pelster}},\ }\href
  {https://doi.org/10.1103/PhysRevB.98.245107} {\bibfield  {journal} {\bibinfo
  {journal} {Phys. Rev. B}\ }\textbf {\bibinfo {volume} {98}},\ \bibinfo
  {pages} {245107} (\bibinfo {year} {2018})}\BibitemShut {NoStop}%
\bibitem [{\citenamefont {{Ohliger}}\ and\ \citenamefont
  {{Pelster}}(2013)}]{Intro_ref_6}%
  \BibitemOpen
  \bibfield  {author} {\bibinfo {author} {\bibfnamefont {M.}~\bibnamefont
  {{Ohliger}}}\ and\ \bibinfo {author} {\bibfnamefont {A.}~\bibnamefont
  {{Pelster}}},\ }\href {https://doi.org/10.4236/wjcmp.2013.32020} {\bibfield
  {journal} {\bibinfo  {journal} {World Journal of Condensed Matter Physics}\
  }\textbf {\bibinfo {volume} {3}},\ \bibinfo {pages} {125} (\bibinfo {year}
  {2013})}\BibitemShut {NoStop}%
\bibitem [{\citenamefont {Anders}\ \emph {et~al.}(2010)\citenamefont {Anders},
  \citenamefont {Gull}, \citenamefont {Pollet}, \citenamefont {Troyer},\ and\
  \citenamefont {Werner}}]{Intro_ref_8}%
  \BibitemOpen
  \bibfield  {author} {\bibinfo {author} {\bibfnamefont {P.}~\bibnamefont
  {Anders}}, \bibinfo {author} {\bibfnamefont {E.}~\bibnamefont {Gull}},
  \bibinfo {author} {\bibfnamefont {L.}~\bibnamefont {Pollet}}, \bibinfo
  {author} {\bibfnamefont {M.}~\bibnamefont {Troyer}},\ and\ \bibinfo {author}
  {\bibfnamefont {P.}~\bibnamefont {Werner}},\ }\href
  {https://doi.org/10.1103/PhysRevLett.105.096402} {\bibfield  {journal}
  {\bibinfo  {journal} {Phys. Rev. Lett.}\ }\textbf {\bibinfo {volume} {105}},\
  \bibinfo {pages} {096402} (\bibinfo {year} {2010})}\BibitemShut {NoStop}%
\bibitem [{\citenamefont {Hu}\ and\ \citenamefont {Tong}(2009)}]{Intro_ref_9}%
  \BibitemOpen
  \bibfield  {author} {\bibinfo {author} {\bibfnamefont {W.-J.}\ \bibnamefont
  {Hu}}\ and\ \bibinfo {author} {\bibfnamefont {N.-H.}\ \bibnamefont {Tong}},\
  }\href {https://doi.org/10.1103/PhysRevB.80.245110} {\bibfield  {journal}
  {\bibinfo  {journal} {Phys. Rev. B}\ }\textbf {\bibinfo {volume} {80}},\
  \bibinfo {pages} {245110} (\bibinfo {year} {2009})}\BibitemShut {NoStop}%
\bibitem [{\citenamefont {Freericks}\ \emph {et~al.}(2009)\citenamefont
  {Freericks}, \citenamefont {Krishnamurthy}, \citenamefont {Kato},
  \citenamefont {Kawashima},\ and\ \citenamefont {Trivedi}}]{Intro_ref_10}%
  \BibitemOpen
  \bibfield  {author} {\bibinfo {author} {\bibfnamefont {J.~K.}\ \bibnamefont
  {Freericks}}, \bibinfo {author} {\bibfnamefont {H.~R.}\ \bibnamefont
  {Krishnamurthy}}, \bibinfo {author} {\bibfnamefont {Y.}~\bibnamefont {Kato}},
  \bibinfo {author} {\bibfnamefont {N.}~\bibnamefont {Kawashima}},\ and\
  \bibinfo {author} {\bibfnamefont {N.}~\bibnamefont {Trivedi}},\ }\href
  {https://doi.org/10.1103/PhysRevA.79.053631} {\bibfield  {journal} {\bibinfo
  {journal} {Phys. Rev. A}\ }\textbf {\bibinfo {volume} {79}},\ \bibinfo
  {pages} {053631} (\bibinfo {year} {2009})}\BibitemShut {NoStop}%
\bibitem [{\citenamefont {Ran\ifmmode~\mbox{\c{c}}\else \c{c}\fi{}on}\ and\
  \citenamefont {Dupuis}(2011)}]{Intro_ref_11}%
  \BibitemOpen
  \bibfield  {author} {\bibinfo {author} {\bibfnamefont {A.}~\bibnamefont
  {Ran\ifmmode~\mbox{\c{c}}\else \c{c}\fi{}on}}\ and\ \bibinfo {author}
  {\bibfnamefont {N.}~\bibnamefont {Dupuis}},\ }\href
  {https://doi.org/10.1103/PhysRevB.84.174513} {\bibfield  {journal} {\bibinfo
  {journal} {Phys. Rev. B}\ }\textbf {\bibinfo {volume} {84}},\ \bibinfo
  {pages} {174513} (\bibinfo {year} {2011})}\BibitemShut {NoStop}%
\bibitem [{\citenamefont {Di~Liberto}\ \emph {et~al.}(2018)\citenamefont
  {Di~Liberto}, \citenamefont {Recati}, \citenamefont {Trivedi}, \citenamefont
  {Carusotto},\ and\ \citenamefont
  {Menotti}}]{GMF_quasiparticle_hole_charcteristics}%
  \BibitemOpen
  \bibfield  {author} {\bibinfo {author} {\bibfnamefont {M.}~\bibnamefont
  {Di~Liberto}}, \bibinfo {author} {\bibfnamefont {A.}~\bibnamefont {Recati}},
  \bibinfo {author} {\bibfnamefont {N.}~\bibnamefont {Trivedi}}, \bibinfo
  {author} {\bibfnamefont {I.}~\bibnamefont {Carusotto}},\ and\ \bibinfo
  {author} {\bibfnamefont {C.}~\bibnamefont {Menotti}},\ }\href
  {https://doi.org/10.1103/PhysRevLett.120.073201} {\bibfield  {journal}
  {\bibinfo  {journal} {Phys. Rev. Lett.}\ }\textbf {\bibinfo {volume} {120}},\
  \bibinfo {pages} {073201} (\bibinfo {year} {2018})}\BibitemShut {NoStop}%
\bibitem [{\citenamefont {Huber}\ \emph {et~al.}(2007)\citenamefont {Huber},
  \citenamefont {Altman}, \citenamefont {B\"uchler},\ and\ \citenamefont
  {Blatter}}]{Intro_ref_13}%
  \BibitemOpen
  \bibfield  {author} {\bibinfo {author} {\bibfnamefont {S.~D.}\ \bibnamefont
  {Huber}}, \bibinfo {author} {\bibfnamefont {E.}~\bibnamefont {Altman}},
  \bibinfo {author} {\bibfnamefont {H.~P.}\ \bibnamefont {B\"uchler}},\ and\
  \bibinfo {author} {\bibfnamefont {G.}~\bibnamefont {Blatter}},\ }\href
  {https://doi.org/10.1103/PhysRevB.75.085106} {\bibfield  {journal} {\bibinfo
  {journal} {Phys. Rev. B}\ }\textbf {\bibinfo {volume} {75}},\ \bibinfo
  {pages} {085106} (\bibinfo {year} {2007})}\BibitemShut {NoStop}%
\bibitem [{\citenamefont {Huber}\ \emph {et~al.}(2008)\citenamefont {Huber},
  \citenamefont {Theiler}, \citenamefont {Altman},\ and\ \citenamefont
  {Blatter}}]{Intro_ref_14}%
  \BibitemOpen
  \bibfield  {author} {\bibinfo {author} {\bibfnamefont {S.~D.}\ \bibnamefont
  {Huber}}, \bibinfo {author} {\bibfnamefont {B.}~\bibnamefont {Theiler}},
  \bibinfo {author} {\bibfnamefont {E.}~\bibnamefont {Altman}},\ and\ \bibinfo
  {author} {\bibfnamefont {G.}~\bibnamefont {Blatter}},\ }\href
  {https://doi.org/10.1103/PhysRevLett.100.050404} {\bibfield  {journal}
  {\bibinfo  {journal} {Phys. Rev. Lett.}\ }\textbf {\bibinfo {volume} {100}},\
  \bibinfo {pages} {050404} (\bibinfo {year} {2008})}\BibitemShut {NoStop}%
\bibitem [{\citenamefont {Altman}\ and\ \citenamefont
  {Auerbach}(2002)}]{Intro_ref_12}%
  \BibitemOpen
  \bibfield  {author} {\bibinfo {author} {\bibfnamefont {E.}~\bibnamefont
  {Altman}}\ and\ \bibinfo {author} {\bibfnamefont {A.}~\bibnamefont
  {Auerbach}},\ }\href {https://doi.org/10.1103/PhysRevLett.89.250404}
  {\bibfield  {journal} {\bibinfo  {journal} {Phys. Rev. Lett.}\ }\textbf
  {\bibinfo {volume} {89}},\ \bibinfo {pages} {250404} (\bibinfo {year}
  {2002})}\BibitemShut {NoStop}%
\bibitem [{\citenamefont {Caleffi}(2022)}]{GMF_Quantum_Fluctuations}%
  \BibitemOpen
  \bibfield  {author} {\bibinfo {author} {\bibfnamefont {F.}~\bibnamefont
  {Caleffi}},\ }\href@noop {} {\bibinfo {title} {Quantum fluctuations beyond
  the gutzwiller approximation in the bose-hubbard model}} (\bibinfo {year}
  {2022}),\ \Eprint {https://arxiv.org/abs/2208.01409} {arXiv:2208.01409}
  \BibitemShut {NoStop}%
\bibitem [{\citenamefont {Nagao}\ and\ \citenamefont
  {Danshita}(2016)}]{Intro_ref_19}%
  \BibitemOpen
  \bibfield  {author} {\bibinfo {author} {\bibfnamefont {K.}~\bibnamefont
  {Nagao}}\ and\ \bibinfo {author} {\bibfnamefont {I.}~\bibnamefont
  {Danshita}},\ }\href {https://doi.org/10.1093/ptep/ptw061} {\bibfield
  {journal} {\bibinfo  {journal} {Progress of Theoretical and Experimental
  Physics}\ }\textbf {\bibinfo {volume} {2016}},\ \bibinfo {pages} {063I01}
  (\bibinfo {year} {2016})}\BibitemShut {NoStop}%
\bibitem [{\citenamefont {Fisher}\ \emph {et~al.}(1989)\citenamefont {Fisher},
  \citenamefont {Weichman}, \citenamefont {Grinstein},\ and\ \citenamefont
  {Fisher}}]{BHM_phase_diagram_1st}%
  \BibitemOpen
  \bibfield  {author} {\bibinfo {author} {\bibfnamefont {M.~P.~A.}\
  \bibnamefont {Fisher}}, \bibinfo {author} {\bibfnamefont {P.~B.}\
  \bibnamefont {Weichman}}, \bibinfo {author} {\bibfnamefont {G.}~\bibnamefont
  {Grinstein}},\ and\ \bibinfo {author} {\bibfnamefont {D.~S.}\ \bibnamefont
  {Fisher}},\ }\href {https://doi.org/10.1103/PhysRevB.40.546} {\bibfield
  {journal} {\bibinfo  {journal} {Phys. Rev. B}\ }\textbf {\bibinfo {volume}
  {40}},\ \bibinfo {pages} {546} (\bibinfo {year} {1989})}\BibitemShut
  {NoStop}%
\bibitem [{\citenamefont {Sengupta}\ and\ \citenamefont
  {Dupuis}(2005)}]{HST_analytical_approximation}%
  \BibitemOpen
  \bibfield  {author} {\bibinfo {author} {\bibfnamefont {K.}~\bibnamefont
  {Sengupta}}\ and\ \bibinfo {author} {\bibfnamefont {N.}~\bibnamefont
  {Dupuis}},\ }\href {https://doi.org/10.1103/PhysRevA.71.033629} {\bibfield
  {journal} {\bibinfo  {journal} {Phys. Rev. A}\ }\textbf {\bibinfo {volume}
  {71}},\ \bibinfo {pages} {033629} (\bibinfo {year} {2005})}\BibitemShut
  {NoStop}%
\bibitem [{\citenamefont {Bradlyn}\ \emph {et~al.}(2009)\citenamefont
  {Bradlyn}, \citenamefont {dos Santos},\ and\ \citenamefont
  {Pelster}}]{Intro_ref_7}%
  \BibitemOpen
  \bibfield  {author} {\bibinfo {author} {\bibfnamefont {B.}~\bibnamefont
  {Bradlyn}}, \bibinfo {author} {\bibfnamefont {F.~E.~A.}\ \bibnamefont {dos
  Santos}},\ and\ \bibinfo {author} {\bibfnamefont {A.}~\bibnamefont
  {Pelster}},\ }\href {https://doi.org/10.1103/PhysRevA.79.013615} {\bibfield
  {journal} {\bibinfo  {journal} {Phys. Rev. A}\ }\textbf {\bibinfo {volume}
  {79}},\ \bibinfo {pages} {013615} (\bibinfo {year} {2009})}\BibitemShut
  {NoStop}%
\bibitem [{\citenamefont {Faccioli}\ and\ \citenamefont
  {Salasnich}(2018)}]{Intro_ref_20}%
  \BibitemOpen
  \bibfield  {author} {\bibinfo {author} {\bibfnamefont {M.}~\bibnamefont
  {Faccioli}}\ and\ \bibinfo {author} {\bibfnamefont {L.}~\bibnamefont
  {Salasnich}},\ }\bibfield  {journal} {\bibinfo  {journal} {Symmetry}\
  }\textbf {\bibinfo {volume} {10}},\ \href
  {https://doi.org/10.3390/sym10040080} {10.3390/sym10040080} (\bibinfo {year}
  {2018})\BibitemShut {NoStop}%
\bibitem [{\citenamefont {Faccioli}\ and\ \citenamefont
  {Salasnich}(2019)}]{Intro_ref_21}%
  \BibitemOpen
  \bibfield  {author} {\bibinfo {author} {\bibfnamefont {M.}~\bibnamefont
  {Faccioli}}\ and\ \bibinfo {author} {\bibfnamefont {L.}~\bibnamefont
  {Salasnich}},\ }\href {https://doi.org/10.1103/PhysRevA.99.023614} {\bibfield
   {journal} {\bibinfo  {journal} {Phys. Rev. A}\ }\textbf {\bibinfo {volume}
  {99}},\ \bibinfo {pages} {023614} (\bibinfo {year} {2019})}\BibitemShut
  {NoStop}%
\bibitem [{\citenamefont {Kobayashi}\ and\ \citenamefont
  {Nitta}(2015)}]{Intro_ref_22}%
  \BibitemOpen
  \bibfield  {author} {\bibinfo {author} {\bibfnamefont {M.}~\bibnamefont
  {Kobayashi}}\ and\ \bibinfo {author} {\bibfnamefont {M.}~\bibnamefont
  {Nitta}},\ }\href {https://doi.org/10.1103/PhysRevD.92.045028} {\bibfield
  {journal} {\bibinfo  {journal} {Phys. Rev. D}\ }\textbf {\bibinfo {volume}
  {92}},\ \bibinfo {pages} {045028} (\bibinfo {year} {2015})}\BibitemShut
  {NoStop}%
\bibitem [{\citenamefont {Gersch}\ and\ \citenamefont
  {Knollman}(1963)}]{Bose-Hubbard_1st}%
  \BibitemOpen
  \bibfield  {author} {\bibinfo {author} {\bibfnamefont {H.~A.}\ \bibnamefont
  {Gersch}}\ and\ \bibinfo {author} {\bibfnamefont {G.~C.}\ \bibnamefont
  {Knollman}},\ }\href {https://doi.org/10.1103/PhysRev.129.959} {\bibfield
  {journal} {\bibinfo  {journal} {Phys. Rev.}\ }\textbf {\bibinfo {volume}
  {129}},\ \bibinfo {pages} {959} (\bibinfo {year} {1963})}\BibitemShut
  {NoStop}%
\bibitem [{aut()}]{author2025supplement}%
  \BibitemOpen
  \href@noop {} {\bibinfo {title} {See supplementary material for ``consistent
  field theory across the mott-insulator to superfluid
  transition''.}}\BibitemShut {Stop}%
\bibitem [{\citenamefont {Bogolyubov}(1947)}]{Bogolyubov_spectrum}%
  \BibitemOpen
  \bibfield  {author} {\bibinfo {author} {\bibfnamefont {N.~N.}\ \bibnamefont
  {Bogolyubov}},\ }\href@noop {} {\bibfield  {journal} {\bibinfo  {journal} {J.
  Phys. (USSR)}\ }\textbf {\bibinfo {volume} {11}},\ \bibinfo {pages} {23}
  (\bibinfo {year} {1947})}\BibitemShut {NoStop}%
\bibitem [{\citenamefont {Pairault}\ \emph {et~al.}(1998)\citenamefont
  {Pairault}, \citenamefont {S\'en\'echal},\ and\ \citenamefont
  {Tremblay}}]{HST_Cumulant_is_bad}%
  \BibitemOpen
  \bibfield  {author} {\bibinfo {author} {\bibfnamefont {S.}~\bibnamefont
  {Pairault}}, \bibinfo {author} {\bibfnamefont {D.}~\bibnamefont
  {S\'en\'echal}},\ and\ \bibinfo {author} {\bibfnamefont {A.-M.~S.}\
  \bibnamefont {Tremblay}},\ }\href
  {https://doi.org/10.1103/PhysRevLett.80.5389} {\bibfield  {journal} {\bibinfo
   {journal} {Phys. Rev. Lett.}\ }\textbf {\bibinfo {volume} {80}},\ \bibinfo
  {pages} {5389} (\bibinfo {year} {1998})}\BibitemShut {NoStop}%
\bibitem [{\citenamefont {Sheshadri}\ \emph {et~al.}(1993)\citenamefont
  {Sheshadri}, \citenamefont {Krishnamurthy}, \citenamefont {Pandit},\ and\
  \citenamefont {Ramakrishnan}}]{HST_MF_equivalence2}%
  \BibitemOpen
  \bibfield  {author} {\bibinfo {author} {\bibfnamefont {K.}~\bibnamefont
  {Sheshadri}}, \bibinfo {author} {\bibfnamefont {H.~R.}\ \bibnamefont
  {Krishnamurthy}}, \bibinfo {author} {\bibfnamefont {R.}~\bibnamefont
  {Pandit}},\ and\ \bibinfo {author} {\bibfnamefont {T.~V.}\ \bibnamefont
  {Ramakrishnan}},\ }\href {https://doi.org/10.1209/0295-5075/22/4/004}
  {\bibfield  {journal} {\bibinfo  {journal} {Europhysics Letters}\ }\textbf
  {\bibinfo {volume} {22}},\ \bibinfo {pages} {257} (\bibinfo {year}
  {1993})}\BibitemShut {NoStop}%
\bibitem [{\citenamefont {Popov}(1988)}]{Popov_book}%
  \BibitemOpen
  \bibfield  {author} {\bibinfo {author} {\bibfnamefont {V.~N.}\ \bibnamefont
  {Popov}},\ }\bibinfo {title} {The modified perturbation scheme for superfluid
  bose systems},\ in\ \href@noop {} {\emph {\bibinfo {booktitle} {Functional
  Integrals and Collective Excitations}}},\ \bibinfo {series and number}
  {Cambridge Monographs on Mathematical Physics}\ (\bibinfo  {publisher}
  {Cambridge University Press},\ \bibinfo {year} {1988})\ p.\ \bibinfo {pages}
  {40–48}\BibitemShut {NoStop}%
\bibitem [{\citenamefont {Rey}\ \emph {et~al.}(2003)\citenamefont {Rey},
  \citenamefont {Burnett}, \citenamefont {Roth}, \citenamefont {Edwards},
  \citenamefont {Williams},\ and\ \citenamefont
  {Clark}}]{Bog_superfluid_fraction}%
  \BibitemOpen
  \bibfield  {author} {\bibinfo {author} {\bibfnamefont {A.~M.}\ \bibnamefont
  {Rey}}, \bibinfo {author} {\bibfnamefont {K.}~\bibnamefont {Burnett}},
  \bibinfo {author} {\bibfnamefont {R.}~\bibnamefont {Roth}}, \bibinfo {author}
  {\bibfnamefont {M.}~\bibnamefont {Edwards}}, \bibinfo {author} {\bibfnamefont
  {C.~J.}\ \bibnamefont {Williams}},\ and\ \bibinfo {author} {\bibfnamefont
  {C.~W.}\ \bibnamefont {Clark}},\ }\href
  {https://doi.org/10.1088/0953-4075/36/5/304} {\bibfield  {journal} {\bibinfo
  {journal} {Journal of Physics B: Atomic, Molecular and Optical Physics}\
  }\textbf {\bibinfo {volume} {36}},\ \bibinfo {pages} {825} (\bibinfo {year}
  {2003})}\BibitemShut {NoStop}%
\bibitem [{\citenamefont {Alves}(2023)}]{Harmonic_Solution}%
  \BibitemOpen
  \bibfield  {author} {\bibinfo {author} {\bibfnamefont {M.~B.}\ \bibnamefont
  {Alves}},\ }\href@noop {} {\bibfield  {journal} {\bibinfo  {journal} {Revista
  Brasileira de Ensino de F{\'\i}sica}\ }\textbf {\bibinfo {volume} {45}},\
  \bibinfo {pages} {e20230152} (\bibinfo {year} {2023})}\BibitemShut {NoStop}%
\end{thebibliography}%

\setcounter{figure}{0}
\setcounter{equation}{0}
\makeatletter
\renewcommand{\thefigure}{S\@arabic\c@figure}
\renewcommand{\theequation}{S\@arabic\c@equation}
\makeatother

\begin{widetext}

\begin{center}
    \Large \textbf{Supplementary Material for ``Consistent Field Theory Across the Mott-Insulator to Superfluid Transition''}\\[1em]
    \normalsize Idan S. Wallerstein and Eytan Grosfeld\\[0.5em]
    \emph{Department of Physics, Ben-Gurion University of the Negev, Beer-Sheva, 84105, Israel}\\[1em]
\end{center}

\section{Failure of the cumulant exansion around the Mott Insulator}
The order parameter that describes the ground state is the solution of the saddle point approximation for the path integral formulation of the partition function, i.e., $\left.\partial_\psi S[\psi,\psi^*]\right|_{\psi=\psi_0}=0$.
Assuming a constant OP $\psi_0$ and using the cumulant expansion, the action is expanded in ${\left| {{\psi _0}} \right|}^2$, due to the $U(1)$ symmetry, where the coefficients are proportional to the $n$-particle correlators of the Bosonic field $\phi$,
\begin{align}
    S\left[ {{\psi _0},{\psi^*_0}} \right]=a_1{\left| {{\psi _0}} \right|^2}\left(\mybraketo{J^{-1}}+\mybraketo{G_1} \right) + \sum\limits_{n = 2}^\infty  {{a_n}{{\left| {{\psi _0}} \right|}^{2n}}\mybraketo{G_n}} .
\end{align}
The coefficients $a_n$ are strictly combinatorial.
The $n$-particle correlators are
\begin{equation}
    {G_n}\left( {{\tau _{1 \leqslant i \leqslant 2n}}} \right) =  - {\left\langle {\left( {\prod\limits_{i = 1}^n {\phi \left( {{\tau _i}} \right)} } \right)\left( {\prod\limits_{i = n + 1}^{2n} {{\phi ^*}\left( {{\tau _i}} \right)} } \right)} \right\rangle _{{\text{loc}}}} =  - \left\langle {\left( {\prod\limits_{i = 1}^n {\phi {e^{\left( {{\tau _i} - {\tau _{i + 1}}} \right){H_{{\text{loc}}}}}}} } \right)\left( {\prod\limits_{i = n + 1}^{2n} {{\phi ^*}{e^{\left( {{\tau _i} - {\tau _{i + 1}}} \right){H_{{\text{loc}}}}}}} } \right)} \right\rangle_{{\text{loc}}} ,
\end{equation}
where the average subscript indicates averaging with respect to the local Hamiltonian
\begin{equation}
    \hat H =  \frac{U}{2}\sum\limits_i^{} {{{\hat n}_i}\left( {{{\hat n}_i} - 1} \right)}  - \mu \sum\limits_i^{} {{{\hat n}_i}},
\end{equation}
and $\mybraketo{G_n}$ is the static component
\begin{align}
\label{eq:G_n static}
   \mybraketo{G_n}=\int\limits_0^\beta  {d\tau_1 \prod\limits_{i = 1}^{2n - 1} {d{(\tau _i-\tau_1)}{G^{\left( n \right)}(\tau_i-\tau_1)}} } .
\end{align}
Using the integral above, in addition to dimensional analysis, $\mybraketo{G_n}$ scales as $\beta \varepsilon^{1-2n}$, where $\varepsilon$ is some energy scale of the local Hamiltonian, i.e., $U$ or $\mu$.
Considering the action for contours of constant $\mu/U=\text{const}$, we only have $U$ as an energy scale, and we can claim that $\mybraketo{G_n}\propto \beta U^{1-2n}$.
This scaling can be shown by direct integration in Eq.~(\ref{eq:G_n static}).
Substituting this scaling in the action, we have
\begin{align}
    S\left[ {{\psi _0},{\psi^*_0}} \right]={\left| {{\psi _0}} \right|^2}\left(b_0+b_1U^{-1}\right) + \sum\limits_{n = 2}^\infty  {{b_n}{{\left| {{\psi _0}} \right|}^{2n}}U^{1-2n}} = U\left[\frac{\left| {{\psi _0}} \right|^2}{U^2}\left(Ub_0+b_1\right) + \sum\limits_{n = 2}^\infty  {{b_n}\frac{{\left| {{\psi _0}} \right|}^{2n}}{U^{2n}}}\right] .
\end{align}
where the $b_0$ depends on $J$, all of the coefficients $b_n$ depend on $\beta$, but none depend on $U$.
The order parameter that minimizes the action is then a function of coefficients $b_n$ and $U$.
For any finite cumulant expansion, one has
\begin{align}
    {\left| {{\psi _0}} \right|^2} = {U^2}F\left( {U{b_0} - {b_1},b_{n\ge2}} \right)\xrightarrow{{U \to 0}}0
\end{align}
The cumulant expansion is therefore not an efficient approximation for the BHM, nonetheless, high order expansions may lead to better results near the phase transition, and even for intermediate values of $zJ/U$ at the cost of computational complexity, but the $U\to0$ limit does require the full calculation of $\ln {\left\langle {\exp \left( {\left[\kern-0.15em\left[ {\psi^* |\phi }  \right]\kern-0.15em\right] + \left[\kern-0.15em\left[ {\phi^* |\psi }  \right]\kern-0.15em\right]} \right)} \right\rangle _{\text{loc}}}$ .

\section{Radius of convergence of the expansion in frequency}
Starting from the Green matrix,
\begin{align}
    {\mathcal{G}}\left( {k,i\omega } \right)&={\left( {\begin{array}{*{20}{c}}
  {J_k^{ - 1} + G\left( {i\omega } \right)}&{{G^\prime }\left( {i\omega } \right)} \\ 
  {{G^\prime }\left( { - i\omega } \right)}&{J_{ - k}^{ - 1} + G\left( { - i\omega } \right)} 
\end{array}} \right)} ,
\end{align}
we write the normal and anomalous Green functions as rational functions
\begin{equation}
\begin{array}{*{20}{l}}
  {\mathcal{G}\left( {k,i\omega } \right)}&{ = \left( {\begin{array}{*{20}{c}}
  {J_k^{ - 1} + \frac{{{G_n}\left( {i\omega } \right)}}{{{G_d}\left( {i\omega } \right)}}}&{\frac{{{{G'}_n}\left( {i\omega } \right)}}{{{{G'}_d}\left( {i\omega } \right)}}} \\ 
  {\frac{{{{G'}_n}\left( { - i\omega } \right)}}{{{{G'}_d}\left( { - i\omega } \right)}}}&{J_{ - k}^{ - 1} + \frac{{{G_n}\left( { - i\omega } \right)}}{{{G_d}\left( { - i\omega } \right)}}} 
\end{array}} \right)} 
\end{array} .
\end{equation}
The spectrum and radius of convergence (singularities) are given by the roots and poles of the determinant, respectively (after the analytic continuation $i\omega\to\omega+0^+$).
\begin{equation}
    \det \left[ {\mathcal{G}\left( {k,\omega } \right)} \right] = \frac{{ - {{G'}_n}\left( { - \omega } \right){{G'}_n}\left( \omega  \right){G_d}\left( { - \omega } \right){G_d}\left( \omega  \right) + {{G'}_d}\left( { - \omega } \right){{G'}_d}\left( \omega  \right)\left( {J_k^{ - 1}{G_d}\left( { - \omega } \right) - {G_n}\left( { - \omega } \right)} \right)\left( {J_k^{ - 1}{G_d}\left( \omega  \right) - {G_n}\left( \omega  \right)} \right)}}{{{{G'}_d}\left( { - \omega } \right){{G'}_d}\left( \omega  \right){G_d}\left( { - \omega } \right){G_d}\left( \omega  \right)}} .
\end{equation}
The poles are dispersionless and are given by the poles of the normal and anomalous Green functions.
Furthermore, the numerator at the limit $J_k\to0$ is identical to the denominator, i.e., the poles equal the roots, and we have $\omega_s=\omega_{k_r}$ for $J_{k_r}=0$.  

\section{Parameters of the low-energy effective theory}
Expanding the Green function gives
\begin{align}
  \mathcal{G}(k,i\omega)\approx \left( {\begin{array}{*{20}{c}}
  {{G} + \frac{{{k^2}}}{2}{{G}^{\left( {2,0} \right)}} + i\omega {{G}^{\left( {0,1} \right)}} - \frac{{{\omega ^2}}}{2}{{G}^{\left( {0,2} \right)}}}&{{G}'- \frac{{{\omega ^2}}}{2}{{G}'}^{\left( 2 \right)}} \\ 
  {{G}' - \frac{{{\omega ^2}}}{2}{{G}'}^{\left( 2 \right)}}&{{G} + \frac{{{k^2}}}{2}{{G}^{\left( {2,0} \right)}} - i\omega {{G}^{\left( {0,1} \right)}} - \frac{{{\omega ^2}}}{2}{{G}^{\left( {0,2} \right)}}} 
\end{array}} \right) ,
\end{align}
where the coefficients ${G}, {G^\prime}$  and their derivatives w.r.t $k$ and $i\omega$ are evaluated at $k=0,i\omega=0$ and are real.
In the MI phase, the Green function is already diagonalized, and the effective action is simply
\begin{align}
   {S^{\left( {{\text{MI}}} \right)}}\left[ {{\varphi _ + },{\varphi _ - }} \right] &= \frac{1}{2}\sum\limits_{k,\omega }^{} {\left[ {G + \frac{{{k^2}}}{2}{G^{\left( {2,0} \right)}} + i\omega {G^{\left( {0,1} \right)}} - \frac{{{\omega ^2}}}{2}{G^{\left( {0,2} \right)}}} \right]{{\left| {{\varphi _{ + ,k}}\left( {i\omega } \right)} \right|}^2}}   \nonumber\\
   &+\frac{1}{2}\sum\limits_{k,\omega }^{} {\left[ {G + \frac{{{k^2}}}{2}{G^{\left( {2,0} \right)}} - i\omega {G^{\left( {0,1} \right)}} - \frac{{{\omega ^2}}}{2}{G^{\left( {0,2} \right)}}} \right]{{\left| {{\varphi _{ - ,k}}\left( {i\omega } \right)} \right|}^2}} .
\end{align}
The approximated spectrum is
\begin{equation}
    {\omega _ \pm } \approx \frac{{ \pm {G^{\left( {0,1} \right)}} + \sqrt {{{\left( {{G^{\left( {0,1} \right)}}} \right)}^2} + 2\left( { - {G^{\left( {0,2} \right)}}} \right)G} }}{{ - {G^{\left( {0,2} \right)}}}} + \frac{{{k^2}{G^{\left( {2,0} \right)}}}}{{2\sqrt {{{\left( {{G^{\left( {0,1} \right)}}} \right)}^2} + 2\left( { - {G^{\left( {0,2} \right)}}} \right)G} }} ,
\end{equation}
which is of the form ${\omega _ \pm } \approx \Delta _ \pm ^{\left( {{\text{MI}}} \right)} + k^2/2 m^{\left( {{\text{MI}}} \right)}$, where $\Delta^{\text{(MI)}}_\pm>0$ and $m^{\text{(MI)}}>0$.

In the SF phase, the action is
\begin{align}
   {S^{\left( {{\text{MI}}} \right)}}\left[ {{\varphi _ + },{\varphi _ - }} \right] &= \frac{1}{2}\sum\limits_{k,\omega }^{} {\left[ {G +G^\prime +\frac{{{k^2}}}{2}{G^{\left( {2,0} \right)}} - \frac{{{\omega ^2}}}{2}\frac{\left(G^{\left( {0,1} \right)}\right)^2+G^\prime\left(G^{\prime(2)}+{G^{\left( {0,2} \right)}}\right)}{G^\prime}} \right]{{\left| {{\varphi _{ + ,k}}\left( {i\omega } \right)} \right|}^2}}   \nonumber\\
   &+\frac{1}{2}\sum\limits_{k,\omega }^{} {\left[ {G -G^\prime +\frac{{{k^2}}}{2}{G^{\left( {2,0} \right)}} + \frac{{{\omega ^2}}}{2}\frac{\left(G^{\left( {0,1} \right)}\right)^2+G^\prime\left(G^{\prime(2)}-{G^{\left( {0,2} \right)}}\right)}{G^\prime}} \right]{{\left| {{\varphi _{ - ,k}}\left( {i\omega } \right)} \right|}^2}} .
\end{align}
In the SF phase, we have $G=G^\prime$ and the action is cast in the form of a Klein-Gordon action,
\begin{align}
\label{eq: excitation action - by band fields}
  {S^{\left( {{\text{SF}}} \right)}}\left[ {{\varphi _ + },{\varphi _ - }} \right] &= \frac{{{\rho_s^{(1)}}}}{2}\sum\limits_{k,\omega}^{} {\left[ {{{\left( {{m_+^{\text{(SF)}} }{c_+}} \right)}^2} + {k^2} + c_+^{ - 2}{\omega ^2}} \right]{{\left| {{\varphi _{+,k}}\left( {i\omega } \right)} \right|}^2}} \nonumber \\
  &+\frac{{{\rho_s^{(1)}}}}{2}\sum\limits_{k,\omega}^{} {\left[{{k^2} + c_-^{ - 2}{\omega ^2}} \right]{{\left| {{\varphi _{-,k}}\left( {i\omega } \right)} \right|}^2}} .
\end{align}
We have, $\rho^{(1)}_s={{G^{\left( {2,0} \right)}}}/2$  for the overall coefficient, the sound velocities are
\begin{equation}
    c_ \pm ^{ - 2} =  \mp \frac{{{{\left( {{G^{(0,1)}}} \right)}^2} \pm G'\left( {{G^{(0,2)}} \pm {{G'}^{\left( 2 \right)}}} \right)}}{{G'{G^{(2,0)}}}} ,
\end{equation}
and the upper band mass is
\begin{equation}
    m_ + ^{\left( {{\text{SF}}} \right)} = \sqrt { - 4\frac{{{{\left( {{G^{(0,1)}}} \right)}^2} + G\left( {{G^{(0,2)}} + {{G'}^{\left( 2 \right)}}} \right)}}{{{{\left( {{G^{(2,0)}}} \right)}^2}}}}>0 .
\end{equation}

The gap of the Klein-Gordon field is the rest mass
\begin{align}
    \Delta _ + ^{\left( {{\text{SF}}} \right)} = m_ + ^{\left( {{\text{SF}}} \right)}c_ + ^{\text{2}} = \sqrt {\frac{{ - 4{G^2}}}{{{{\left( {{G^{(0,1)}}} \right)}^2} + G\left( {{G^{(0,2)}} + {{G'}^{\left( 2 \right)}}} \right)}}}>0.
\end{align}

\section{Pure amplitude and phase Hamiltonian}

\begin{figure*}
\centering
    \includegraphics[width=.48\linewidth]{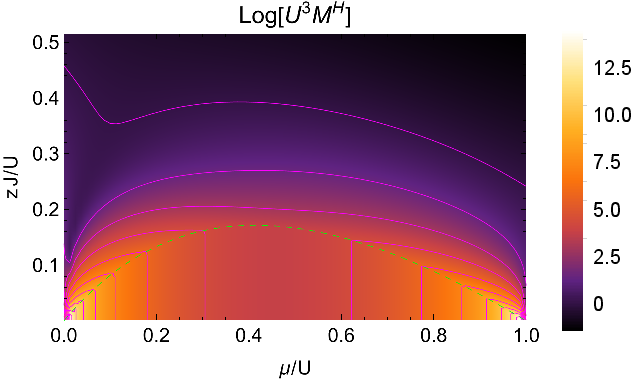}
    \hspace{0.1cm}
    \includegraphics[width=.48\linewidth]{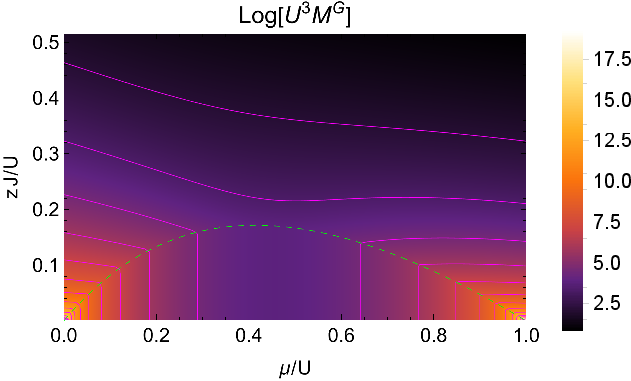}
\caption{
Dimensionless parameters for the low energy effective Hamiltonian, contour lines in magenta, green dashed line represents the MI-SF phase transition. 
(a) Mass of the amplitude mode, (b) Mass of phase mode.
Masses are equal in the MI phase.}
\label{fig:Masses}
\end{figure*}
Starting from the Euclidean Lagrangian density
\begin{align}
    &\mathcal{L}\left[ {h\left( {{\mathbf{r}},\tau } \right),g\left( {{\mathbf{r}},\tau } \right)} \right] = \left( {G + G'} \right){h^2} + \left( {G - G'} \right){g^2} + \frac{1}{2}{G^{\left( {2,0} \right)}}\left( {{{\left( {\nabla h} \right)}^2} + {{\left( {\nabla g} \right)}^2}} \right)\hfill \nonumber\\
    & + i{G^{\left( {0,1} \right)}}\left( {g{\partial _\tau }h - h{\partial _\tau }g} \right) - \frac{1}{2}{\left( {{\partial _\tau }h} \right)^2}\left( {{G^{\left( {0,2} \right)}} + {{G'}^{\left( 2 \right)}}} \right) - \frac{1}{2}{\left( {{\partial _\tau }g} \right)^2}\left( {{G^{\left( {0,2} \right)}} - {{G'}^{\left( 2 \right)}}} \right) ,
\end{align}
we write the real-time Lagrangian density $\mathcal{L}(\tau)\to-\mathcal{L}(i t)$
\begin{align}
  &\mathcal{L}\left[ {h\left( {{\mathbf{r}},t} \right),g\left( {{\mathbf{r}},t} \right)} \right] =  - \left( {G + G'} \right){h^2} - \left( {G - G'} \right){g^2} - \frac{1}{2}{G^{\left( {2,0} \right)}}\left( {{{\left( {\nabla h} \right)}^2} + {{\left( {\nabla g} \right)}^2}} \right)\hfill \nonumber \\
  &  - {G^{\left( {0,1} \right)}}\left( {g{\partial _t}h - h{\partial _t}g} \right) - \frac{1}{2}{\left( {{\partial _t}h} \right)^2}\left( {{G^{\left( {0,2} \right)}} + {{G'}^{\left( 2 \right)}}} \right) - \frac{1}{2}{\left( {{\partial _t}g} \right)^2}\left( {{G^{\left( {0,2} \right)}} - {{G'}^{\left( 2 \right)}}} \right) .
\end{align}
The canonical momenta are
\begin{align}
  {p_h} &= {\partial _{{\partial _t}h}}\mathcal{L} =  - {G^{\left( {0,1} \right)}}g - \left( {{G^{\left( {0,2} \right)}} + {{G'}^{\left( 2 \right)}}} \right){\partial _t}h ,\hfill \\
  {p_g} &= {\partial _{{\partial _t}g}}\mathcal{L} =  + {G^{\left( {0,1} \right)}}h - \left( {{G^{\left( {0,2} \right)}} - {{G'}^{\left( 2 \right)}}} \right){\partial _t}g .
\end{align}
The Legendre transform leads to the Hamiltonian density
\begin{align}
  \mathcal{H} &= \frac{{{{\left( {{p_h} + {G^{\left( {0,1} \right)}}g} \right)}^2}}}{{ - 2\left( {{G^{\left( {0,2} \right)}} + {{G'}^{\left( 2 \right)}}} \right)}} + \left( {G + G'} \right){h^2} \hfill  + \frac{1}{2}{G^{\left( {2,0} \right)}}{{{\left( {\nabla h} \right)}^2}} \nonumber\\
   & + \frac{{{{\left( {{p_g} - {G^{\left( {0,1} \right)}}h} \right)}^2}}}{{ - 2\left( {{G^{\left( {0,2} \right)}} - {{G'}^{\left( 2 \right)}}} \right)}} + \left( {G - G'} \right){g^2} + \frac{1}{2}{G^{\left( {2,0} \right)}}{{{\left( {\nabla g} \right)}^2}} .
\end{align}
As a final step, we take the Fourier transform to find the Hamiltonian density in momentum space
\begin{align}
\label{eq:LEFH}
    {{\mathcal{H}}^+_k} = \frac{{{{\left| {{p_{h,k}} + {G^{\left( {0,1} \right)}}{g_k}} \right|}^2}}}{{ - 2\left( {{G^{\left( {0,2} \right)}} + {{G'}^{\left( 2 \right)}}} \right)}} + \frac{{{{\left| {{p_{g,k}} - {G^{\left( {0,1} \right)}}{h_k}} \right|}^2}}}{{ - 2\left( {{G^{\left( {0,2} \right)}} - {{G'}^{\left( 2 \right)}}} \right)}} +\left( {G + G' + \frac{{{k^2}}}{2}{G^{\left( {2,0} \right)}}} \right){\left| {{h_k}} \right|^2} + \left( {G - G' + \frac{{{k^2}}}{2}{G^{\left( {2,0} \right)}}} \right){\left| {{g_k}} \right|^2} .
\end{align}
This Hamiltonian is in the form of an anisotropic two-dimensional harmonic oscillator with a uniform magnetic field (Fig.~\ref{fig:Masses})
\begin{align}
    H = \frac{{\left( {{p_h} + {Bg/2}} \right)}^2}{2M_h} + \frac{{\left( {{p_g} - {Bh/2}} \right)}^2}{2M_g} +\frac{K_hh^2}{2} +\frac{K_gg^2}{2} , 
\end{align}
where we drop the $k$ dependency. The Hamilton equation of motion is 
\begin{equation}
    \frac{d}{{dt}}\left( {\begin{array}{*{20}{c}}
  h \\ 
  g \\ 
  {{p_h}} \\ 
  {{p_g}} 
\end{array}} \right) = \left( {\begin{array}{*{20}{c}}
  0&{\frac{B}{{2{M_h}}}}&{\frac{1}{{{M_h}}}}&0 \\ 
  { - \frac{B}{{2{M_g}}}}&0&0&{\frac{1}{{{M_g}}}} \\ 
  { - \frac{{{B^2}}}{{4{M_g}}} - {K_h}}&0&0&{\frac{B}{{2{M_g}}}} \\ 
  0&{ - \frac{{{B^2}}}{{4{M_h}}} - {K_g}}&{ - \frac{B}{{2{M_h}}}}&0 
\end{array}} \right)\left( {\begin{array}{*{20}{c}}
  h \\ 
  g \\ 
  {{p_h}} \\ 
  {{p_g}} 
\end{array}} \right) .
\end{equation}
The eigenvectors of this equation define new variables $a_+$, $a_-$, and their complex conjugate, such that their time evolution is harmonic with the excitations spectrum frequency, i.e., $a_\pm(t)=a_\pm(0)\exp(-i\omega_\pm t)$.
Defining the harmonic and cyclotron frequencies, ${\omega _{0,i}(k)} = \sqrt {{K_i(k)}/{M_i}}$  and ${\omega _{c,i}} = B/{M_i}$, respectively, these frequencies are
\begin{equation}
    \omega _ \pm ^2 = \frac{1}{2}\left[ {{\omega _{c,h}}{\omega _{c,g}} + \omega _{0,h}^2 + \omega _{0,g}^2 \pm \sqrt {{{\left( {{\omega _{c,h}}{\omega _{c,g}} + \omega _{0,h}^2 + \omega _{0,g}^2} \right)}^2} - 4\omega _{0,h}^2\omega _{0,g}^2} } \right],
\end{equation}
which agree with the low-energy effective theory spectrum to the $k^2$ order.

\end{widetext}

\end{document}